\documentstyle[12pt]{article}
\newcommand{\LL}{\log( \frac{\Lambda^2}{M_W^2} )}
\newcommand{\LH}{\log( \frac{M_H^2}{M_W^2} )}

\begin{document}
\begin{titlepage}
%\begin{center}
%\Large {EUROPEAN ORGANIZATION FOR NUCLEAR RESEARCH \\}
%\end{center}
\vspace*{0.1cm}
\rightline{\large CERN-TH 6670}
\rightline{\large LPTHE-Orsay 92/56}
\rightline{\large FTUAM-92/34}
\vspace*{0.4cm}
\vspace*{0.1cm}
%\rightline{\today}
\vspace*{0.2cm}
\begin{center}
\Huge{\bf One Loop Effects of Non-Standard Triple Gauge Boson Vertices \\}
\vspace*{2cm}
\large{\bf P. Hern\'andez$^{(1)}$ and F.J. Vegas$^{(2)}$ \\}
\vspace*{0.4cm}
\end{center}
\begin{center}
$^{(1)}$ { CERN-CH-1211 Geneva 23, Switzerland\\
                      and\\
           Universidad Aut\'onoma de Madrid\\
           Cantoblanco, Madrid 28049, Spain\\}
$^{(2)}$ { Laboratorie de Physique Theorique et Hautes Energies\\
  Universit\'e de Paris-sud, Bat. 211, Orsay 91405, France}
\end{center}
\vspace{1.1cm}
\begin{center}
ABSTRACT
\end{center}
Low energy effects of generic extensions of the Standard Model
can be comprehensively parametrized in terms of higher dimensional
effective operators. After the success of all the recent precission tests
on the Standard Model, we argue that any
sensible description of these extensions at the Z-scale
must be stable under higher order quantum corrections. The imposition
of $SU(2)_L \times U(1)_Y$ gauge invariance seems to be the simplest
and most natural way to fulfill this requirement.
With this assumption, all the possible deviations
from the standard triple gauge boson vertices can be consistently
parametrized in terms of a finite set of gauge invariant
operators. We deal here with those operators that do not give any
tree level effect on present experimental observables and
constrain them by computing their effects at the
one-loop level. We conclude that for
a light Higgs boson, the direct measurement at LEP200 can improve
present bounds on these "blind directions", while for a heavy Higgs
it is most unlikely to provide any new information.

\vspace{0.1cm}
\vfill
\leftline{\large CERN-TH 6670}
\leftline{\large LPTHE-Orsay 92/56}
\leftline{\large FTUAM-92/34}
\leftline{December 1992}
\vspace*{0.2cm}
\end{titlepage}
\newpage
\section{Introduction}  \label{intro}
Half a dozen of the Standard Model (SM) predictions are already tested to first
order in their quantum corrections\footnote{ Quantitatively
only the running of QED and QCD couplings has been significatively
tested up to now, although including also the proper weak corrections
one obtains a better agreement with experimental data.
We acknowledge discussions on this point with
L. Maiani and L. Okun.}.
 For these quantum corrections to be finite,
subtle cancellations occur, which require the algebraic relations imposed
by the theory's gauge invariance on the various couplings to be exactly
satisfied.
In this situation, the natural hypothesis is to assume
that theories beyond the SM must be such that they
generate an effective theory, which
preserves at least the gauge symmetry $SU(2) \times U(1)$ over the symmetry
breaking scale  $v$.
Other assumptions need to prove their "quantum consistency".
As emphasized in
\cite{deru1,deru3},
this has often been overlooked in the literature, leading
to overly optimistic expectations concerning the "new physics" sensitivity
of future machines.

The leading low energy effects of a generic
"meta-theory" at a scale $\Lambda \,(\geq v)$ can be
comprehensively and systematically
parametrized in terms of a linear combination of higher dimensional
operators, constructed out of the light fields. If the
low energy limit of this "meta-theory" has the symmetries and light fields
of the minimal SM (it takes care of its own ultraviolet
divergences), the
higher dimensional operators must also preserve those symmetries.
Besides, even
if the low energy effective theory at the Z-scale is not the
renormalizable linear SM (this is for instance the case of a theory with a
strongly interacting symmetry breaking sector), the successes of the
minimal SM still imply that $\Lambda$ cannot be independent of $v$ and
that the
pattern of the symmetry breaking must be $SU(2)_L \times SU(2)_C
\rightarrow SU(2)$, at least\footnote{This pattern is
not the most general \cite{chano}, but it doesn't seem very sensible
at present to adopt the point of view of Burguess and London
\cite{london}
of substituting the hypothesis of a general symmetry principle
($SU(2)_C$) by an unnatural set of fine tunnings.} to a precission of
1\%.

In \cite{deru1} the effects of "new-physics" on
the structure and strength of the triple gauge-boson vertices (TGVs)
were systematically analysed in an effective Lagrangian parametrization.
The basis of this method is discussed in sections 2 and 3.
It was found that all the operators affecting the TGVs can
be expressed as a linear combination of six independent ones, which were
chosen to be those giving tree level effects on present observables.
All the rest can be expressed in terms of these by means
of the equations of motion. However, some combinations of the basis
operators are such that their tree level effects on present observables
exactly cancel. These are the so called "blind directions".
Even when there is no known symmetry or dynamical
reason why a meta-theory would be so contrived as to generate a low
effective Lagrangian pointing exclusively in these directions, in order
to leave out any theoretical
prejudice, we do not exclude this possibility. In this paper, we constrain
them directly from present data through
their one-loop effects.
In section 4, we present the results of our computation, which has also
been partially carried out in \cite{chori}.
In section 5, we
compare LEP-1 and LEP-2 sensitivities. We conclude in section 6
that the better
chances for LEP-2 concerning an anomalous correction to the triple
gauge boson
vertices in these directions, come from a relatively light Higgs
(as expected also for many other reasons), while for a heavy Higgs LEP-1
constraints are already considerably better.

\section{ Effective Lagrangian Parametrization}
Recently, there has been some controversy in relation with the "uses and
abuses" of Effective Lagrangian parametrizations in the search for non-
standard effects in present and future experiments \cite{reac}.
Questions like what are the symmetries one must impose on an Effective
Lagrangian or whether the latter can be used in higher orders of perturbation
theory without loosing their predictive power, seem to be misunderstood
even though the principles of this approach were settled long ago
\cite{fermi,classics}.
We briefly review the ideas behind the
Effective Lagrangian parametrization of New Physics and how they can
consistently be used in higher orders of perturbation theory.

 The Operator Product Expansion \cite{wilson} is an
example of the factorization property of renormalizable theories,
by which, at a given energy scale, the effects on physical
observables of the much higher energy modes, can be factorized in
the effective couplings of the light ones.
In any renormalizable theory in which the typical scale of
some fields (given for example by their mass) becomes large,
the effect of this heavy sector on Feynman's amplitudes between light
states
can be expanded in perturbation theory as a sum of local operators
of the light fields (multiplied by the appropiate dimensional
coefficients which depend on the heavy scale)\footnote{
Some care is needed when considering a theory in which masses are generated
by spontaneous symmetry breaking, for in this case masses are given by some
dimensionless couplings $\lambda$ times the vacuum expectation value (vev)
of a scalar
field. When the mass is large due to a large Yukawa coupling, while
the vev is light, the large mass expansion does not exist in general
\cite{kazama}.}.

Suppose that at very high energy we have a renormalizable theory
in  which the masses of some fields are much higher than the energy
of our experiment (if the masses are generated by spontaneous symmetry
breaking we assume that the vev for that sector is large).
If the light sector is symmetric under a given gauge subgroup
(then it is renormalizable),
obviously the full theory must also be symmetric and necessarily
the operators of the large mass expansion also
preserve this gauge symmetry\footnote{In fact the
operators must be BRS-invariant, so in principle, there could be
operators involving ghost fields. However, as shown in \cite{weinb}
the Faddeev-Popov procedure can be done in two steps, firstly for
the "heavy subgroup" when integrating the heavy sector
and secondly for the "light" subgroup when defining the effective theory.
This way no extra interactions appear in the "light" ghost sector.}.
In neglecting the contributions vanishing as $\Lambda
\rightarrow \infty$, we are left with local operators
of light fields of dimension $d\leq4$, whose coefficients may
contain positive powers of $\Lambda$ times logarithms. However, as long
as they preserve the simmetries of the light sector, they just
produce a finite renormalization of the light
couplings and all the non-vanishing dependence on  $\Lambda$ is
physically unobservable. The heavy
fields decouple \cite{appelcar}. This is in fact the case
in many physically
interesting situations where the new physics comes from extended gauge
groups or larger symmetries. We will refer to this situation as the
decoupling case.

An effective Lagrangian parametrization is natural here,
because the renormalizability of the light theory provides
a power counting which controls the non-renormalizable interactions
coming from the heavy sector. If we want to consider the
effects up to some given order in 1/$\Lambda$,  we can truncate
the expansion at that
order and the theory so obtained takes care of its own ultraviolet
divergences \cite{polchi}.

Thus, in parametrizing as generally as possible the leading effects of
any decoupling new physics,
we will consider a combination
of light operators of the smallest possible dimension greater than 4,
\newpage
which preserve the gauge symmetries of the light sector:

\begin{eqnarray}
{\cal L}_{eff} = {\cal L}_{light} + \sum_{j} {\alpha}_{j}
\frac {O_j^{d_j}}{\Lambda^{d_j-4}}.
\end{eqnarray}

This is the most general large mass expansion of a renormalizable
meta-theory whose low energy limit is ${\cal L}_{light}$. The quantum
corrections of this lagrangian are well defined order by order in
$1/\Lambda$, which implies, in particular, that
there is no physical divergence coming out at the one loop level.

On the contrary, if the light theory is non-renormalizable, there must
be some physical cutoff related to the heavy scale.
The effects of the heavy fields cannot decouple, as they must
compensate the tower of divergent light counterterms needed to make
the full theory finite. Even though these effects can also be written as
a sum of local operators of the light fields, this is in general of no
use since there are now contributions with positive
powers of $\Lambda$ that can not be renormalized in ${\cal L}_{light}$
and, a priori, it is not clear that there
is any useful expansion to sum the leading effects in $\Lambda$,
in terms of a finite set of couplings.
The quantum corrections in this case contain non-renormalizable
divergences and, contrary to what is claimed in \cite{london},
this has a clear physical meaning: the
cutoff can not be sent to infinity, furthermore, there must be some
natural relation between the light and heavy scales.
One example of the fact that these divergences
are physical is the Minimal Standard Model. Suppose that
an elementary Higgs do exist, but its mass is larger than $M_Z$
and $M_W$. Then, at LEP energies, we can integrate it out and work
in the resulting effective theory, which is a non-linear sigma model.
The non-renormalizable divergences that appear in the perturbative
computation of quantum corrections in the effective theory, turn
out to be the corrections in $M_H$ computed in the
full renormalizable theory \cite{appel}, they do not disappear!
In fact, it is through these divergences that one is able to estimate
the natural
scale of the couplings in the effective theory (contrary to the case
of a decoupling sector, for which there is no naturality argument to
estimate $\Lambda$ and it can only be determined experimentally).

\section{ Anomalous Triple Gauge Boson Vertices}
The literature abounds in estimates of the LEP-2 sensitivity  to
the structure
of the gauge boson vertex, based on a general Lorentz invariant
parametrization of that vertex which does not preserve the gauge
symmetry \cite{chorisa}. As we have seen, the effect of these new
interactions is to break the renormalizability
of the light theory (as the gauge symmetry is lost) and consequently its
predictivity. This leads to very optimistic expectations concerning
the "new physics" sensitivity of future machines, but
makes it difficult to understand how it can be that
half a dozen of the predictions of the SM are already succesfully tested
to first order in their quantum corrections \cite{deru2}.

 In \cite{deru1} a general Effective Lagrangian parametrization of novel
effects in the structure of the TGVs was considered.
Two main possibilities were analysed:

1) The case of Decoupling type of New Physics. In this case the
hypothesis is that the Minimal Standard Model is the correct theory
to describe physical phenomena at the Z scale. Novel effects would then
come from extra particles, larger gauge symmetries, compositeness...
characterized by a mass scale $\Lambda$ distinct from the scale $v$. In
particular, this means that the unspecified heavy objects are assumed not
to acquire their masses from the standard machinery of symmetry breaking,
though they may well be involved in the mechanisms that trigger it.
The leading effects were shown to come from d=6 operators, which by the
previous reasoning must be SU(2)xU(1) invariant. In \cite{deru1}, the basis
operators and the blind direction $O_W \equiv i \vec{W^{\nu}_{\mu}}
\times \vec{W^{\lambda}_{\nu}} \cdot \vec{W^{\mu}_{\lambda}}$
were studied in detail. Here we deal
with the remaining blind-directions. In the decoupling case, we have
two more of them:

\begin{eqnarray}
O_{B\Phi} & \equiv &  i B^{\mu\nu} {( D_{\mu} \Phi )}^{\dag}
D_{\nu} \Phi,\\
O_{W\Phi} & \equiv &  i \vec{W^{\mu\nu}} {( D_{\mu} \Phi )}^{\dag}
\vec{\sigma} D_{\nu} \Phi.
\end{eqnarray}
where $W^a_{\mu\nu} \equiv  \partial_{\mu} W_{\nu}^a -
\partial_{\nu} W_{\mu}^a - g \epsilon_{abc} W_{\mu}^b W_{\nu}^c$;
and $B_{\mu\nu} \equiv \partial_{\mu} B_{\nu} - \partial_{\nu}
B_{\mu}$.

2) The case of a strongly interacting symmetry breaking sector
does not fit in the previous picture, because the
light theory is in this case a non-linear sigma model
which is not perturbatively renormalizable. We assume that the
interactions responsible for the generation of intermediate vector boson
masses have a global $SU(2)_L \times SU(2)_C$ symmetry, with $SU(2)_C$
 the accidental
custodial symmetry \cite{silvi} of the standard potential of scalar
doublets.
This is the only natural situation given the experimental fact that $\rho
\simeq 1$ to a precision of a percent.  Without any further assumption,
the most general Lagrangian is that of a nonlinearly realized $SU(2)_L
\times SU(2)_C$ which breaks to $SU(2)$. After switching on
the gauge fields, this implies the $SU(2)_L \times U(1)$ gauge symmetry
\cite{coli}.

  As shown by Weinberg \cite{wein},
a loop expansion in this theory is equivalent to a momentum expansion,
so that only a finite number of operators are needed to describe the
physical phenomena at low energies (this number increasing with
the order in the momentum expansion).
There is a natural dimension-full parameter which suppresses these
non-renormalizable terms ($\Lambda = 4 \pi v$) \cite{georgi}
and the effective Lagrangian is just a Taylor expansion in
$U$ , $\frac{D}{\Lambda}$, $\frac{W_{\mu\nu}}{\Lambda^2}$ and
$\frac{B_{\mu\nu}}{\Lambda^2}$, where $U$ is the unitary matrix
describing the longitudinal gauge-boson degrees of freedom,
transforming as a $(2,2)$ of $SU(2)_L \times SU(2)_C$. The covariant
derivative acting on $U$ is the usual one:

\begin{eqnarray}
D_{\mu} U \equiv \partial_{\mu} U + i g \frac{\vec{\sigma}}{2}
\vec{W_{\mu}} U - i g U \frac{\sigma_3}{2} B_{\mu}.
\end{eqnarray}
It is important to remark that in the gauge sector, the loop
expansion is not necessarily a low-energy expansion.

The leading  non-standard effects on TGVs come from operators of
$d_{\chi}$=4. After reducing the list with the use of the equations
of motion, we are
left \cite{deru1} with three independent ones and three blind directions.
Here, we will consider the effect of these blind directions\footnote{
In Gasser-Leutwyler notation \cite{gasser}, the $L_9$ operator corresponds
to a combination of these operators $(\beta_2 = \beta_3 )$.}

\begin{eqnarray}
{\cal L}_2  & \equiv & i g' \beta_2 B^{\mu\nu} Tr \{ T [ (D_{\mu}U)
U^{\dag},(D_{\nu}U)U^{\dag}] \},\\
{\cal L}_3 & \equiv & i g \beta_3 Tr \{ \vec{W_{\mu\nu}} \vec{\sigma}
[(D_{\mu}U) U^{\dag},(D_{\nu}U)U^{\dag}] \},\\
{\cal L}_9 & \equiv & i g \beta_9 Tr \{ T W^{\mu\nu}\} Tr \{
T [ (D_{\mu}U)U^{\dag},(D_{\nu}U)U^{\dag}] \}.
\end{eqnarray}
where $T \equiv U \sigma_3 U^{\dag}$. Those operators containing
$T$ are not custodial preserving.
Even when we assume that the symmetry breaking
sector is symmetric under $SU(2)_C$, the coupling to hypercharge
breaks this symmetry and operators containing $T$ can appear, although
they always have a factor $g'$.

It turns out that in practice,         the effects of these operators are
proportional to the effects of equivalent operators in the linear realization
with the identification $M_H \sim \Lambda$ (the linear model with a Higgs
scalar is a regulator for the nonlinear model \cite{appel}):

\begin{eqnarray}
{\cal L}_2 & \Rightarrow & \frac{8 g' \beta_2}{v^2} O_{B\Phi}{|}_{(\Phi
\rightarrow v)},\\
{\cal L}_3 & \Rightarrow & \frac{8 g \beta_3}{v^2} O_{W\Phi}{|}_{(\Phi
\rightarrow v)},\\
{\cal L}_9 & \Rightarrow & \frac{-32 i g \beta_9}{v^4} ( \Phi^{\dag} W^{\mu\nu}
\Phi)(D_\mu \Phi)^{\dag}D_\nu\Phi {|}_{(\Phi \rightarrow v)}.
\end{eqnarray}

We see that the first two operators on the right-hand side are two of
the three blind directions of the decoupling case, while the third one is
an operator of d=8, which was not leading in that case. This shows
the different power counting of the two parametrizations.

In fact, at any given dimension, one expects more operators in the
non-decoupling case. If we consider for example the Standard Model with
a higgs in the intermediate region (not so heavy as to render
the theory non-renormalizable, but heavier than the energies of LEP-1
and LEP-2 experiments), we can work in both the linear and
non-linear realizations. When we integrate the higgs out, all the
contributions that appear in the linear case as finite corrections
in $M_H$, should appear after its integration as new effective operators.
The linear treatment would be more adequate then, because there are
fewer degrees of freedom.

\section{Present Data Constraints}

Following \cite{deru1}, we will use $\alpha$, $G_F$  and $M_z$
as the input parameters for the minimal Standard Model, because they
are the most accurately measured quantities.
Then, for constraining the previous
 operators from present data
we will use as observables the W mass, the leptonic and hadronic widths of
the Z, the forward-backward asymmetry in leptonic Z decays, the $\tau$
polarization $P_\tau$
and the ratio of inclusive neutral to charged-current neutrino cross
sections on aproximately isoscalar targets $R_\nu$:

\begin{eqnarray}
M_W & = & 80.13 \pm 0.31 GeV \cite{mw} \nonumber \\
\Gamma_l & = & 83.52 \pm 0.33 MeV \cite{nash} \nonumber \\
\Gamma_h & = & 1742 \pm 8 MeV  \cite{nash} \nonumber \\
A_{FB}^l & = & 0.0157 \pm 0.003  \cite{nash} \nonumber \\
P_\tau & = & -.140 \pm 0.024  \cite{nash} \nonumber \\
R_\nu & = & 0.308 \pm 0.002 \cite{rnu}
\end{eqnarray}

To the experimental error
in $R_\nu$ we have added a theoretical error of $1 \%$ to reflect the
theoretical
uncertainty associated with the charm threshold in the charged currents,
discussed and estimated in \cite{26}. We adopt the safe recipe of adding
linearly the theoretical uncertainties to the experimental errors. We
shall choose to perform our analysis at the 2 $\sigma$ level, the inputs
to our limits on novel effects are then twice the quoted errors.

None of these observables are affected at tree level by the insertion of
blind operators, their effects start at the one-loop level. Our
calculation has been done in a generic $\xi$-gauge, and we have been
able to check the cancellation of the
$\xi$-dependence
in physical observables. We have used two regularization procedures,
dimensional regularization and a simple momentum cutoff. Both give
the same results with the identification\footnote{A
common misconception is the statement that there are no quadratic
divergences in dimensional regularization. There are, and they correspond
to poles at d=2.}
 of quadratic divergences
with poles at $d=2$ and logarithmic ones with poles at $d=4$.
As discussed before, in the decoupling case all
divergences but logarithms must get renormalized in the couplings of
the SM, so they must also cancel in physical observables. The coefficients
of the logarithmic divergences                             are the
coefficients of the renormalization group logarithms that appear in
the running of the effective Lagrangian  coefficients from
the scale $\Lambda$ to the
Z-scale. Our computation is valid up to these logarithms, not including
constant terms.
In the nonlinear case, however, the leading contribution comes from
quadratic divergences (poles in $d=2$) that do not cancel. These clearly
correspond to
the terms in $M_H^2$ of the linear case.
There are also some leading contributions coming from finite terms
in $M_H^2$.

 We express the effect of the insertion of blind operators in terms of
the following dimensionless parameters:

\begin{eqnarray}
{\delta}_{B \Phi} & \equiv & \frac {g s}{c} \frac{{M_W}^2}
{(4 \pi)^2 \Lambda^2}
{\alpha}_{B \Phi}, \\
{\delta}_{W{\Phi}} & \equiv & \frac {g s^2}{c^2} \frac {{M_W}^2}
{(4\pi)^2{\Lambda}^2}
{\alpha}_{W\Phi} , \\
\delta_9 & \equiv & -8 \frac{s^2}{c^2} \frac{g^4 \beta_9}{(4 \pi)^2}.
\end{eqnarray}
where, from now on, $s=\sin {\theta_W}$ and $c=\cos {\theta_W}$.

The quantum effects of these operators appear either as
boson self-energies $\Pi^{\gamma\gamma}$, $\Pi^{\gamma Z}$, $\Pi^{ZZ}$
and
$\Pi^{WW}$,
or as corrections to the $Z f f$ vertex ($\delta c_L^f$) and
the $W l \nu$ vertex ($\delta g_{Wl\nu}$).
We collect all this quantum corrections in the appendix A.
In terms of these objects, the shifts induced in the renormalization
parameters are:

\begin{eqnarray}
\frac {\Delta\alpha}{\alpha} & = & \frac{{\Pi}_{\gamma\gamma}}
{q^2} {|}_{q^2 = 0} ,\\
\frac {\Delta{M_Z}^2}{{M_Z}^2} & = & \frac{{\Pi}_{ZZ}(q^2)}{{M_Z}^2}
{|}_{q^2 = {M_Z}^2} ,\\
\frac {\Delta G_F}{G_F} & = & [\frac{2\delta g_{Wl\nu}}{g_{Wl\nu}}
- \frac {{\Pi}_{WW}(q^2)}{{M_W}^2} ] {|}_{q^2 = 0}.
\end{eqnarray}

All the physical observables can now be expressed in terms of the
preceeding objects.
\newpage
We parametrize the shifts in the widths in terms
of $\delta \gamma_f$, $\delta \kappa_f$,
which also contain the main
contribution from the standard radiative corrections
\footnote{The radiatively-corrected
standard predictions on $\Gamma_f$ can be cast in the form: \newline
$\Gamma_f \simeq \frac{G_F M_Z^3}{3 \pi \sqrt{2}} N_c (1 + \delta
\gamma_f) ([c_f^R]^2+[c_f^L]^2)$,
with $c_f^R = -(1+\delta \kappa_f) Q_f sin^2 \theta_Z$ and
$c_f^L = T_3^f + c_f^R$.}:

\begin{eqnarray}
\delta{\gamma}_f & = & -\frac{\Delta G_F}{G_F}- \frac{\Delta {M_Z}^2}
{{M_Z}^2} + Re( \frac{{\Pi}_{ZZ}(q^2)-{\Pi}_{ZZ}({M_Z}^2)}{q^2-{M_Z}^2})
+ 2 \frac{ \delta {c_L}^f}{({c_L}^f-{c_R}^f)},
\end{eqnarray}

\begin{eqnarray}
\delta{\kappa}_f  & = & - \frac{c^2}{(c^2-s^2)}(\frac{\Delta\alpha}
{\alpha}-\frac{\Delta{M_Z}^2}{{M_Z}^2}-\frac{\Delta G_F}{G_F})
- \frac {c}{s} \frac{Re({\Pi}_{\gamma Z} (q^2))}{q^2}
- \frac{ \delta {c_L}^f}{({c_L}^f-{c_R}^f)}.
\end{eqnarray}

The observables we will use to constrain the blind operators are:

\begin{eqnarray}
\frac {\delta {M_W}^2}{{M_W}^2} & = & \frac{{\Pi}_{WW}({M_W}^2)}{{M_W}^2}
- \frac{\Delta {M_Z}^2}{{M_Z}^2}
+ \frac{s^2}{(c^2-s^2)}(\frac{\Delta\alpha}
{\alpha}-\frac{\Delta {M_Z}^2}{{M_Z}^2}-\frac{\Delta G_F}{G_F}),
\end{eqnarray}

\begin{eqnarray}
\frac{\delta{\Gamma}_l}{\Gamma_l} & = & \delta{\gamma}_l(M_Z^2)-0.250
\delta{\kappa}_l (M_Z^2),
\end{eqnarray}

\begin{eqnarray}
\frac{\delta{\Gamma}_h}{\Gamma_h} & = & \delta{\gamma}_q (M_Z^2)-0.318
\delta{\kappa}_q (M_Z^2),
\end{eqnarray}

\begin{eqnarray}
\frac{\delta {A_{FB}^l}}{A_{FB}^l} & = & 4 \frac{s^2 \delta \kappa^l}
{g_v} \;
\frac{{g_a}^2-g_v^2}{g_a^2+g_v^2},
\end{eqnarray}

\begin{eqnarray}
\frac{\delta P_{\tau}}{P_{\tau}}  & = & -2 \; \frac{s^2
\delta \kappa^l}{g_v} \; \frac{(g_a - g_v)^2}{g_v ^2 + g_a ^2},
\end{eqnarray}

\begin{eqnarray}
\frac{\delta R_{\nu}}{R_{\nu}} & = & \delta \gamma_q (0) + \delta
\gamma_\nu (0) + 2
[ \frac{ \delta \kappa_u(0)(c_L^u c_R^u  + \frac{1}{3} {c_R^u}^2)
+ \delta \kappa_d(0) (c_L^d c_R^d + \frac{1}{3} {c_R^d}^2) }{
{c_L^u}^2 +{c_L^d}^2 +\frac{1}{3} ( {c_R^u}^2 + {c_R^d}^2)}].
\end{eqnarray}

The standard one-loop contributions to these observables
depend sensitively on $m_t$ and weakly on $M_H$. In \cite{deru1} the
only blind direction considered did not give any extra effect on $M_H$ and
the uncertainty due to the dependence on this parameter was summed linearly
with
the experimental error. Our operators, on the contrary, have a strong
dependence
on $M_H$ at the one loop level,
so we will consider separately the cases of a light Higgs ($M_H = 50
GeV$)
and a heavy Higgs ($M_H \sim 1 TeV$).

We assume no acccidental cancellations between the contributions of the
various operators to the different observables and extract from existing
experiments the combined allowed domains in the ($m_t$,${\delta}_i$)
planes, whose projections are the 95.5 \% confidence level (2 $\sigma$)
intervals
on the individual variables, for the limiting values of $M_H$.
The bounds we find are weaker for a light Higgs and become more restrictive
as $M_H$ grows, due to the quadratic dependence
on $M_H$ of the loop effects. For $O_{B\Phi}$, the constraints become
more restrictive as $M_H$ increases in  the whole region from
$M_H = 50$ GeV to 1 TeV, while for $O_{W\Phi}$
this behaviour starts only after $M_H \simeq 260$ GeV. In the region
between 50 and 260 GeV the constraints are weaker for a heavier Higgs due
to cancellations between terms in $M_H^2$ and the other quantum
corrections.

\begin{itemize}
\item{Case of a light Higgs boson }

In the following table we show the $2 \sigma$ contraints on
$\delta_{B\Phi}$ and $\delta_{W\Phi}$, as a function of $M_H$
for any value of $m_t$:

\begin{tabular}{|l|c|c|} \hline
$M_H = 50 GeV$  &
$-1.5 \cdot {10}^{-4} \leq \delta_{B\Phi} \leq 3.7 \cdot {10}^{-4}$&
$-1.8 \cdot {10}^{-4} \leq \delta_{W\Phi} \leq 3.8 \cdot 10^{-4}$\\
  \hline
$M_H = 100 GeV$  &
$-1.6 \cdot {10}^{-4} \leq \delta_{B\Phi} \leq 3.0 \cdot {10}^{-4}$&
$-2.0 \cdot {10}^{-4} \leq \delta_{W\Phi} \leq 4.6 \cdot {10}^{-4}$\\
  \hline
$M_H = 260 GeV$  &
$-8.0 \cdot {10}^{-5} \leq \delta_{B\Phi} \leq 1.8 \cdot {10}^{-4}$&
$-2.8 \cdot {10}^{-4} \leq \delta_{W\Phi} \leq 1.38 \cdot {10}^{-3}$\\
  \hline
$M_H = 500 GeV$  &
$-4.0 \cdot {10}^{-5} \leq \delta_{B\Phi} \leq 1.1 \cdot {10}^{-4}$&
$-1.4 \cdot {10}^{-4} \leq \delta_{W\Phi} \leq 4.9 \cdot {10}^{-4}$\\
  \hline
\end{tabular}

These blind operators also generate new couplings involving scalars
not present at tree level
in the standard model, like the $Z H_0 \gamma$ vertex.
We will translate the present experimental limits on the decay
$Z \rightarrow H_0 \gamma$ into new constraints on the $\delta 's$ for
a light Higgs ($M_H \leq M_Z$). Let $A= A_0 + A_{B \Phi} +A_{W \Phi}$
be the amplitude for $Z \rightarrow H_0\gamma$ decay. The corresponding
width is

\begin{eqnarray}
\Gamma(Z \rightarrow H_0 \gamma) = |A|^2 \frac{E^3_{\gamma}}{12 \pi}.
\end{eqnarray}

The standard $A_0$ is dominated by the triangle graph with intermediate W's
and is equal to \cite{50} :

\begin{eqnarray}
A_0 \simeq - \frac{e \alpha}{4 \pi sin^2 \theta M_W} [ 4.56+0.25
(\frac{M_H}{M_W})^2],
\end{eqnarray}

practically independent of $M_H$ for $M_H \leq M_Z$. With the same
normalization:

\begin{eqnarray}
A_{B\Phi} = - \frac{\alpha_{B\Phi} M_W}{\Lambda^2} ,\\
A_{W\Phi} =  \frac{\alpha_{W\Phi} M_W}{\Lambda^2}.
\end{eqnarray}

The ratio $\Gamma(Z \rightarrow H_0\gamma)/\Gamma_0(Z \rightarrow H_0\gamma
)$ varies from $\sim 0$ to $\sim 4.$ in the interval $|\delta_{B(W)\Phi}|
\leq 3.8 \cdot 10^{-5}$, indicating a strong sensitivity to
the new physics. The trouble
is that current bounds on the branching ratio B for $Z \rightarrow H_0
\gamma$ are not overly restrictive. From L3 result \cite{L3} $B \leq 10^{-
3}$
(for $48 \leq M_H \leq 86 GeV$) we get the following constraints for
$M_H = 50 GeV$:

\begin{eqnarray}
|\delta_{B\Phi}| \leq 2.4 \cdot 10^{-4},\\
|\delta_{W\Phi}| \leq 2.4 \cdot 10 ^{-4}.
\end{eqnarray}

It is remarkable that these constraints, which are completely independent
of the preceeding ones,  are of the same
order of magnitude and do not depend on $m_t$. Also the dependence on
$M_H$ is weaker (of  course, only inside the range
$M_H \leq M_Z$).

\item {Case of a heavy Higgs ($M_H \sim 3 TeV$)}

In the case of a heavy Higgs, we take $\Lambda \simeq M_H \simeq 4 \pi v
\; (\sim 3 TeV)$, which corresponds to the natural situation in the
non-linear case. For any value of $m_t$:

\begin{eqnarray}
-9.4 \cdot 10^{-3} & \leq \beta_2 \leq & 2.2 \cdot 10^{-2},\\
-1.5 \cdot 10^{-2} & \leq \beta_3 \leq & 3.9 \cdot 10^{-2},\\
-1.1 \cdot 10^{-2} & \leq \beta_9 \leq & 4.7 \cdot 10^{-3}.
\end{eqnarray}

These latter constraints are much more restrictive due to the existence
of quadratic divergences.
Obviously, these leading contributions can be
renormalized in the couplings of some other non-blind operators of the same
dimension. However, for this chiral expansion to be natural,
we expect that the divergent part coming from the loop contribution
is (but for additional powers of the couplings $g$ or $g'$)
of the same order of
the  renormalized coupling \cite{appel}.
 Thus,
we constrain these chiral operators indirectly by constraining the
counterterms they necessarily generate, which are
non-blind to present observables. On naturality grounds, we expect these
results to be a correct estimation of the order of magnitude. It is
straightforward to
check that the constraints on $\beta 's$ we have just derived
are
typically a factor $g^2$ worse than the constraints obtained in
\cite{deru1} for non-blind operators.

\end{itemize}

In \cite{chori}, only the situation
$\delta_{B\Phi} = \delta_{W\Phi} = \delta$ was studied. This
corresponds to the effect of the operator $L_9$ in Gasser-Leutwyler's notation
(GL).
In order to compare our numerical results with theirs, we have also
obtained the bounds form present data in this situation. We consider the
case of a light Higgs ($M_H = 60 GeV$ and $\Lambda = 300 GeV$):

\begin{eqnarray}
-2.4 \cdot {10}^{-4} \leq \delta \leq 8.0 \cdot {10}^{-4},
\end{eqnarray}
while their constraints translate into
\begin{eqnarray}
-3.0 \cdot {10}^{-4} \leq \delta \leq 1.5 \cdot {10}^{-3}.
\end{eqnarray}

We find similar results for other values of $M_H$ and $\Lambda$.
To our understanding the differences between our results and those in
\cite{chori} come from the fact that they use Altarelli-Barbieri
$\epsilon's$ as present data constraints instead of directly measured
observables, with the unavoidable propagation of errors.
Besides, they partially lose the advantage of the different dependence
on $m_t$ of the single observables.

In a recent reference \cite{vanderbij}, it is claimed that there are
quartic contributions to $\delta \rho$ for all the operators except
those that break $SU(2)_C$ via a minimal coupling to hypercharge
\cite{87}. These quartic divergences appear at next order in the
coupling (they are $\sim {\delta}^2 $), as expected from simple
power counting \cite{appel}, and there is no physical
reason to impose their cancellation on naturality grounds. Rather,
if one considers at this order all the counterterms of lower chiral
dimension, these quartic divergences disappear from physical observables.
Besides, the minimal coupling to hypercharge is not even realized
in the standard model with a heavy Higgs boson. Operators with non
minimal coupling do appear \cite{appel} in this case, although they
always contain a factor $g'$.

\section{LEP200 Sensitivity}

Now, we turn to study the sentitivity of the future LEP-200 experiment to these
operators. There, $\sqrt{s} \simeq 200GeV$ and the channel $e^{+}e_{-}
\rightarrow W^{+}W_{-}$ is opened, so any anomaly in the self-coupling
of vector bosons will contribute at the tree level.
We use the standard notation for the trilinear couplings:

\begin{eqnarray}
{\cal L}_0^{(3)}(V)= -  i e g_V [( W_{\mu\nu}^{\dag} W^{\mu} -
W_{\mu\nu} W^{\dag \mu}) V^{\nu} + \kappa_V W_{\mu}^{\dag} W_{\nu}
V^{\mu\nu}] - i e g_V \frac{\lambda_V}{{M_W}^2}[V^{\mu\nu}
W^{\dag}_{\nu\rho} W^{\rho}_{\mu}],
\end{eqnarray}
where $W_{\mu\nu} = \partial_{\mu} W_{\nu} - \partial_{\nu} W_{\mu}$
and V = $ \gamma$, Z. Blind operators produce tree level shifts on
the couplings $\kappa_V$ and $g_V$.

A very sensitive direct test concerns the differential cross section
$ d \sigma / d cos \theta_+$, with $\theta_+$  the
$e^+ W^+$ scattering angle. The possible non-standard effects asssociated
with our blind directions will come through shifts in the quantities
$\kappa_Z ,\,\kappa_\gamma,\,g_Z,\,g_\gamma,\,M_W,\,c_L^e$ and $c_R^e$.
For LEP-200 we will consider only the tree-level shifts and neglect
the one-loop effects in $M_W$, $c_L^e$ and $c_R^e$ as well as the
standard radiative corrections.

\begin{eqnarray}
\delta \kappa_\gamma & = & \lambda_{B\Phi} + \lambda_{W\Phi} -\frac{1}{2}
 \lambda_9 ,\\
\delta \kappa_Z & = &  - \frac{s^2}{c^2} ( \lambda_{B\Phi} + \lambda_{W\Phi} )
- \frac{1}{2} \lambda_9 ,\\
\delta g_\gamma & = & 0 ,\; \; \; \; \; \\
\delta g_Z & = & \frac{1}{c^2} \lambda_{W \Phi} g_Z.
\end{eqnarray}

where we have defined:

\begin{eqnarray}
\lambda_{B \Phi} \equiv \frac{g c}{4 s}
\frac{v^2 \alpha_{B\Phi}}{\Lambda^2},\;  \; \; \; \; \;
\lambda_{W \Phi} \equiv \frac{g}{4}
\frac{v^2 \alpha_{W \Phi}}{\Lambda^2},\; \; \; \; \; \; \;
\lambda_9 \equiv -8 g^2 \beta_9.
\end{eqnarray}

We have Monte-Carlo generated $10^4$ W-pairs at $\sqrt{s} = 200 GeV$,
which is a generous estimation of LEP2 statistics, and performed
 $\chi^2$ tests of significance of deviations from the standard
differential cros-section at various values of $\lambda's$. The errors
considered are only statistical.

In Figs.1,
 we show the biggest allowed domains on the planes $(m_t, \delta )$
( at $M_H = 50GeV$ for $O_{B\Phi}$ and at $M_H = 260GeV$ for $O_{W\Phi}$)
together with the expected LEP200 constraints, for
$\Lambda = 1 TeV$. For the operator $O_{B\Phi}$, LEP200 will improve
the present upper bound in a factor 2-3, at most. However, the
sensitivity of LEP200 to $O_{W\Phi}$ is $\sim 5$ times better,
and consequently there is an improvement of an order of magnitude with
respect to present bounds in this case.

As we explained before, the worst constraints from present data on this
operator are obtained for $M_H \sim 260 GeV$, where
the allowed domain in the $(m_t, \delta_{W\Phi})$ plane
extends to very large values of $m_t$.
In fact,
the upper limit on $\delta_{W\Phi}$ grows more than a factor 4
from the one obtained at $M_H = 50 GeV$, but it
can only be saturated  if $m_t$ turns out to be
$\sim 350 GeV$. In this situation, the bound for $\delta_{W\Phi}$ from
present data can be read from Fig. 1:

\begin{eqnarray}
 -2.8\cdot 10^{-4} \leq \delta_{W\Phi} \leq 1.38 \cdot 10^{-3},
\end{eqnarray}
to be compared with the expected sensitivity of LEP200, also shown in
Fig. 1:
\begin{eqnarray}
 -3.7 \cdot 10^{-5} \leq \delta_{W\Phi} \leq 3.1 \cdot 10^{-5}.
\end{eqnarray}

If $m_t$ is lighter than $200 GeV$, the present upper
bound would go down to $\delta_{W\Phi} \leq 4.0 \cdot 10^{-4}$ and, for
greater values, present data constrain  $\delta_{W\Phi}$ to be in a
 band of width $\sim 4 \cdot 10^{-4}$ whose central value grows
 linearly with $m_t^2$. The allowed domain from LEP200
slightly intersects this region and, in this sense, future constraints
will complement present ones and not supersede them (giving a new
bound on the top mass in case LEP200 fails to detect a
non-standard TGV).

The results for the case of a heavy Higgs are gathered in Figs. 2.
 They
show the comparison between the $\chi^2$-test limits from LEP200 and
present constraints for the three operators of (8)-(10).
These results are very similar to that of
non-blind operators \cite{deru1}. In particular, the
bounds on $\beta_2 (\sim L_9^R)$ and $\beta_3 (\sim L_9^L)$ at LEP1,
 are much better than those
that can be obtained from their tree level effects in LEP200 and
CDF, although not competitive with those from
SSC and LHC \cite{valencia}\footnote{In terms of $L_9^R$ and $L_9^L$,
our constraints translate into
 $-13.9    \leq  \L_9^R \leq 6.0$ and
 $-24.6    \leq  \L_9^L \leq 9.5$.}.

 As we have explained in the previous section,
we constrain these operators indirectly by constraining the
counterterms they generate, which are non-blind to present observables.
In other words, present constraints imply that the sensitivity of
the experiments LEP200 and CDF to the TGVs will
not be enough to measure a value of $L_9$ coupling of the order which is
natural for a strongly interacting symmetry breaking sector \cite{espriu}.

It is interesting to study whether
a rise in energy to $E_{cm} = 500 GeV$ at NLC, will
give much better constraints compared to LEP200's. We have generated
25000 events at $E_b = 250 GeV$
(which corresponds to an integrated luminosity of $10 fb^{-1}$)
and the bounds we obtain for the $\beta's$ are the
following:

\begin{eqnarray}
-1.8 \cdot 10^{-2}  \leq \beta_2 \leq 7.2 \cdot 10^{-3} \\
-7.2 \cdot 10^{-3}  \leq \beta_3 \leq 5.5  \cdot 10^{-2}.
\end{eqnarray}

NLC sensitivity to the blind directions is a factor $\sim 5$
better than that of LEP200, and competitive with present
constraints in the case of a heavy Higgs.
Obviously, the information obtained
from direct measurement has less uncertainty than that obtained
from the loop effects, so if nature has chosen any
of these blind directions
to "deform" the standard TGV's, NLC will certainly improve
our present knowledge.

\section{Conclusions}

 Our computation of the one-loop effects of the "blind operators"
completes the analysis started in \cite{deru1} of
the bounds on  non-standard triple
gauge boson vertices from present experimental data.
We have argued, yet once again, the necessity of imposing the gauge
symmetry on the effective Lagragian, for this approach to be
stable under higher order perturbations. Present experiments are
already sensitive to radiative corrections, consequently, any meaningful
search for possible new physics by means of an
effective Lagrangian parametrization must be such that these
corrections are well defined.

We have considered two main possibilities for departures from
the original version of the standard model. The first is that
in which the effective theory at the Z-scale is correctly described
by the minimal standard model with a relatively light Higgs
and the new physics appears as larger symmetries or extended gauge
groups. Secondly, we also considered the case of
an spontaneous symmetry breaking
sector involving some sort of strongly coupled dynamics, where the
elementary scalars may play no role at all. If the Higgs is not found
within the range up to $O(1 TeV)$, this possibility
seems more likely.

The one-loop corrections due to these operators depend quadratically
on $M_H$ (or equivalently, on the cutoff in the case of a strongly
interacting symmetry breaking sector). We have studied separately
the limiting cases of a heavy and a light Higgs. We conclude that
LEP200 is more sensitive to any new physics pointing
in these blind directions if the Higgs is light:
for $O_{B\Phi}$, present bounds
are only a factor 2-3 worse than our conservative estimation
of future LEP200 sensitivity, while for $O_{W\Phi}$ this factor grows
to an order of magnitude.

On the contrary,
in the case of a heavy Higgs and considering the natural situation
$M_H \sim 4 \pi v$, present constraints are already considerably better
than those that can be obtained in LEP200. On naturality grounds, we
have argued that the quadratic dependence on the cutoff is physically
relevant in the non-linear realization, implying that the size of
the counterterm must be that of the
quadratic terms in $\Lambda$, within an order of magnitude.

Even though it is not natural to expect that an extension of the standard
model will point exclusively in these blind directions,
present data already constrain considerably this possibility and,
still in those situations where
present bounds are weaker, the allowed regions on the planes $(m_t,
\delta's)$ from present and future experiments are independent
to a large extent.

Although our estimation of LEP200 sensitivity is quite optimistic,
this analysis can certainly be refined.
The measurement of helicity amplitudes,
for instance, is expected to increase the sensitivity by a factor 2.
Also, the measurement of $M_W$ with much better precission, can
improve the indirect constraints.

Finally we have also estimated the sensitivity of NLC to these blind
directions and found an increase of a factor $\sim 5$ with respect
to LEP200.

\section{Acknowledgments}
We thank \'Alvaro De R\'ujula,
Andy Cohen, Bel\'en Gavela, Juanjo G\'omez Cadenas, Takeo Inami,
Elisabeth Jenkins,
Aneesh Manohar, Eduard Masso,
Olivier P\'ene and Juan Terr\'on for encouragement
and/or illuminating discussions. We are indebted to \'Alvaro De R\'ujula,
Bel\'en Gavela and Olivier P\'ene for careful reading of the manuscript.
We also acknowledge the Ministerio de Educaci\'on y Ciencia and the
Universidad Aut\'onoma de Madrid for financial
support during the completion of this work.

\newpage
\section{Appendix A}
\begin{itemize}

\item{Self-energies:}
\end{itemize}

\begin{eqnarray}
\Pi_{\gamma\gamma} & = & - 2 c^2 ( {\delta}_{B \Phi}
+ {\delta}_{W{\Phi}} - \frac{1}{2}{\delta_9} )
  [ {\Lambda}^2- (\frac {q^2}{6} + 3 {M_W}^2)\LL ] \frac{q^2}{M_W^2}
\end{eqnarray}
\begin{eqnarray}
\Pi_{\gamma Z}     & = & \frac {c}{s}  [ ( s^2 {\delta}_{B{\Phi}}
- c^2 {\delta}_{W{\Phi}} + \frac{c^2-s^2}{4} \delta_9)
(2 \frac {\Lambda^2}{{M_W}^2}
- (\frac {q^2}{3{M_W}^2} + 6)\LL ) \nonumber \\
&   & + [ \frac {9}{4}( {\delta}_{B{\Phi}}
+ {\delta}_{W{\Phi}} -\frac{1}{2} \delta_9 )
- \frac {s^2}{4 c^2}( {\delta}_{B{\Phi}}
- {\delta}_{W{\Phi}}-\frac{1}{2}
\delta_9) - \frac {3}{2} \xi {\delta}_{W{\Phi}}]\LL] q^2  \nonumber\\
&   & + \frac{c}{s} ( \delta_{W\Phi}
-\delta_{B\Phi} +\frac{\delta_9}{2}) \frac{M_H^2}{4 M_W^2}
( \LH -\frac{1}{2} ) q^2
\end{eqnarray}
\begin{eqnarray}
\Pi_{ZZ}     & = & ( \frac{c^4}{s^2} \delta_{W{\Phi}}
+ s^2 \delta_{B{\Phi}} + \frac{c^2}{2} \delta_9 )
( - 2 \Lambda^2 + (q^2/3)\LL ) \frac{q^2}{M_W^2}  \nonumber \\
& + & ( \frac{c^2}{s^2} {\delta}_{W{\Phi}}
+ {\delta}_{B{\Phi}} + \frac{c^2}{2 s^2} \delta_9)
[ [  q^2 ( \frac{1}{6c^2}+ 1-6c^2- \frac{M_H^2}{2 M_W^2}) \nonumber\\
& + & \frac{3}{2c^2} ( M_H^2 +M_Z^2-\frac{q^2}{3} )] \LL
+ \frac{1}{2} (q^2- 3 M_Z^2) \frac{M_H^2}{M_W^2}
( \LH - \frac{1}{2} ) +\frac{3}{c^2} \Lambda^2 ] \nonumber\\
& + &  \frac{c^2}{s^2} \delta_{W\Phi} [ q^2
(3 +12 c^2 - \frac{2}{3 c^2} - 3 \xi ) + ( 3 (\xi+1) M_Z^2
- \frac{q^2}{c^2})] \LL -\frac{1}{ s^2} \delta_9 \Lambda^2
\end{eqnarray}
\begin{eqnarray}
\Pi_{WW} & = & ( \delta_{B{\Phi}} + \frac {c^2}{s^2}
\delta_{W{\Phi}})
[ - \frac{q^2}{3} + 3 ( \Lambda^2 + (\frac {M_Z^2}{2}
-\frac {3}{2} M_W^2 - \frac {q^2}{6})\LL) ]  \nonumber \\
& + &  \frac {c^2}{s^2} \delta_{W{\Phi}}
[ [ ((M_W^2 ( \frac {1}{2c^2}+ \frac {28}{3}- 3 \xi) + \frac{q^2}{3}
- \frac {M_H^2}{2})\LL-2 {\Lambda}^2) \frac{q^2}{M_W^2} \nonumber\\
& + & 3 ( (\xi+6){M_W}^2- \frac {3}{2}{M_Z}^2
+ \frac {{M_H}^2}{2})\LL]  + \frac{1}{2} ( q^2 - 3 M_W^2 )
\frac{M_H^2}{M_W^2} ( \LH - \frac{1}{2} ) ] \nonumber\\
& + &  \frac{c^2}{s^2} \delta_9 [
 - \frac{\Lambda^2}{2} +
( \frac{3q^2}{4} \xi -\frac{13q^2}{12}
 +\frac{q^2}{4}
+ \frac{M_Z^2}{4} +\frac{5}{4} M_W^2- \frac{3}{4} \xi M_W^2)\LL)]
\end{eqnarray}

\begin{itemize}
\item{Vertex corrections:}
\end{itemize}

\begin{eqnarray}
\delta {c_L}^f & = & \frac {3}{2} \frac{c^2}{s^2} (\xi+1)
( {c_L}^f-{c_R}^f) \delta_{W\Phi} \LL
\end{eqnarray}

\begin{eqnarray}
\frac{\delta g_{Wl \nu}}{g_{Wl \nu}} & = & \frac {3}{4} \frac {c^2}{s^2}
(\xi+1) [(\delta_{W \Phi}- \frac{\delta_9}{2})
({c_L}^{\nu}-{c_L}^l+c_R^f) + \frac{1}{c^2} \delta_{W \Phi}
({c_L}^{\nu}-{c_L}^l)] \LL
\end{eqnarray}

\newpage
\begin{center}
\Large { Figure Captions }
\end{center}

\vspace*{0.7 cm}
Figs. 1 Allowed $2 \sigma$ contours in the $(\delta_i, m_t)$ planes
from present data and $M_H = 50 GeV$/$M_H = 260GeV$ for $O_{B\Phi}$/
$O_{W\Phi}$.
The dashed domains subtends the
values of $\delta_i's$ that can not be distinguished from zero at
LEP200 at the $2 \sigma$ level.

\vspace*{0.7 cm}
Figs. 2  $\chi^2$ test fo significance of the effect of $\lambda_i's
\neq 0$ on $d \sigma /d cos \theta_+$. The horizontal line shows the
$2 \sigma$ sensitivity for $10^4$ W-pairs at $\sqrt{s} = 200 GeV$,
the projections along the vertical arrows delimit the interval
of $\lambda_i's$  inside which a LEP-2 measurement would test the
hypothesis $\lambda_i's \neq 0$ with less than $2 \sigma$ significance.
Vertical bands encompasses the values fo $\lambda_i's$ currently allowed
by the lower-energy tests, for $M_H \sim 4 \pi v$.
\end{document}